\documentclass[sigplan,screen,nonacm]{acmart}

\AtBeginDocument{%
  }

\setcopyright{none}
\copyrightyear{2027}
\acmYear{2027}
\usepackage{microtype}
\usepackage{algorithm}
\usepackage{algpseudocode}
\usepackage{wrapfig}
\usepackage{xparse}
\usepackage{placeins}
\usepackage[export]{adjustbox}
\usepackage{listings}
\usepackage{mathpartir}
\usepackage{tcolorbox}
\usepackage[normalem]{ulem}
\usepackage{enumitem}
\lstdefinestyle{pseudo}{
    basicstyle=\ttfamily\small,
    keywordstyle=\bfseries,
    morekeywords={loop, if, else},
    xleftmargin=1mm,
}
\lstdefinestyle{pseudosmall}{
    basicstyle=\ttfamily\scriptsize,
    keywordstyle=\bfseries,
    morekeywords={loop, if, else},
    xleftmargin=1mm,
}
\usepackage{cleveref}
\crefformat{section}{§#2#1#3}
\Crefformat{section}{§#2#1#3}
\crefrangeformat{section}{§§#3#1#4--#5#2#6}
\Crefrangeformat{section}{§§#3#1#4--#5#2#6}
\crefmultiformat{section}{§§#2#1#3}{, #2#1#3}{, #2#1#3}{, #2#1#3}
\Crefmultiformat{section}{§§#2#1#3}{, #2#1#3}{, #2#1#3}{, #2#1#3}
\crefformat{equation}{(#2#1#3)}
\Crefformat{equation}{(#2#1#3)}
\crefrangeformat{equation}{(#3#1#4)--(#5#2#6)}
\Crefrangeformat{equation}{(#3#1#4)--(#5#2#6)}
\crefmultiformat{equation}{(#2#1#3)}{ and~(#2#1#3)}{, (#2#1#3)}{, and~(#2#1#3)}
\Crefmultiformat{equation}{(#2#1#3)}{ and~(#2#1#3)}{, (#2#1#3)}{, and~(#2#1#3)}

\newcommand{\appdetail}[1]{#1}

\usepackage{xspace}
\usepackage{url}
\usepackage{listings}
\usepackage{xcolor}
\usepackage{verbatimbox}
\usepackage{makecell}
\usepackage{multirow}
\usepackage{multicol}
\usepackage{pgfplots}
\usepackage{pgfplotstable}
\usepackage{filecontents}
\usepackage{soul}
\usepackage{cleveref}
\usepackage{tikz}
\usetikzlibrary{calc,fit,positioning,decorations.pathreplacing,arrows.meta}
\usepackage{booktabs}

\newcommand{\rc}{R}

\newcommand{\instrOp}{\mathit{instrOp}}
\newcommand{\guardFn}{\mathit{guard}}
\newcommand{\ctrlOp}{\mathit{controlOp}}

\newcommand{\prd}{\mathit{producer}}         
\newcommand{\cns}{\mathit{consumer}}         
\newcommand{\shared}{S}                      
\newcommand{\parscope}{P}                    
\newcommand{\isGT}{\mathsf{isGT}}                

\begin{document}

\title{A Barrier-Free Synchronization Algorithm for Multi-Engine AI Accelerators}


\author{Chungha Sung}
\authornote{Both authors contributed equally to this research.}
\affiliation{%
\institution{Amazon, Inc.}
  \country{USA}
}
\email{chunghs@amazon.com}

\author{Nikil V. Shyamsunder}
\authornotemark[1]
\authornote{Work done while at Amazon.}
\affiliation{%
  \institution{Cornell University}
  \country{USA}
}
\email{nvs26@cornell.edu}

\author{Hanliang Zhang}
\affiliation{%
  \institution{Amazon, Inc.}
  \country{USA}
}
\email{hlz@amazon.com}

\author{Daniel Kroening}
\affiliation{%
\institution{Amazon, Inc.}
  \country{USA}
}
\email{dkr@amazon.com}

\author{Joonwon Choi}
\affiliation{%
\institution{Amazon, Inc.}
  \country{USA}
}
\email{jwch@amazon.com}

\renewcommand{\shortauthors}{Sung, Shyamsunder, Zhang, Kroening, and Choi}

\ccsdesc[500]{Software and its engineering~Compilers}
\ccsdesc[500]{Theory of computation~Program verification}
\ccsdesc[300]{Computer systems organization~Parallel architectures}

\keywords{semaphore allocation, multi-engine synchronization, barrier elimination, bisimulation, formal verification, AI accelerator, Lean proof assistant}


\begin{abstract}

Multi-engine AI accelerators such as AWS Trainium comprise specialized compute engines that execute in parallel, and the compiler must synchronize the data dependencies between them.
For straight-line code this is simple: each dependency reduces to waiting for a \emph{threshold} count of instruction completions, which the compiler computes statically.
Loops admit no such static threshold; a simple solution inserts \emph{all-engine barriers} at iteration boundaries, resetting synchronization state so each loop body can be treated as straight-line, at the cost of parallelism.

We present a \emph{barrier-free} synchronization algorithm that instead enforces each dependency precisely across structured control flow with arbitrarily nested, dynamically bounded loops.
The key idea is to compute dynamic thresholds at runtime from tracked loop iteration counts.

We implemented it as a compiler backend pass at the AWS Neuron ISA level.
%
On a suite of ML kernels, it reduces latency 10--45\% relative to the barrier-based baseline, achieves a $3.3\times$ speedup on a synchronization-bound microbenchmark, and often matches or exceeds hand-tuned manual allocation.

Issuing a consumer too early violates its dependency, while issuing too late unnecessarily stalls execution. We formally characterize the minimum synchronization required for correctness and verify in the Lean proof assistant, via bisimulation, that our algorithm meets this criterion.

\end{abstract}

\maketitle

\section{Introduction}

\label{sec:intro}


AWS Trainium~\cite{trn1,trn2,trn3} is a family of AI accelerators comprising multiple specialized compute engines that share on-chip buffers and execute concurrently~\cite{nki-arch}.
Each engine runs its own instruction stream asynchronously, and within an engine instructions are deeply pipelined, so a later instruction may begin before an earlier one completes~\cite{nki-arch}.
Data dependencies therefore arise both \emph{across} engines (e.g., a DMA load must complete before the Tensor Engine consumes the data) and \emph{within} an engine.
AWS's Neuron compiler~\cite{neuron-compiler} analyzes dependencies and enforces them through hardware \emph{semaphores}: shared atomic counters incremented when an instruction completes and tested against a threshold before issue.
For straight-line code, each engine's semaphore counts instruction completions, and a dependent instruction waits for it to reach a statically computed threshold---its \emph{wait value}.
This is simple, efficient, and correct, but only applies to acyclic code.


Loops break this static-threshold scheme: the same instruction completes many times, so a fixed ordinal threshold no longer identifies a single completion.
Unrolling would restore it at the cost of code size, but this requires statically known trip counts, infeasible for modern ML workloads with dynamically bounded loops (e.g., variable-length attention, Mixture-of-Experts).
A simple alternative synchronizes all engines at every loop entry and loop iteration boundary via all-engine barriers: each engine drains its in-flight instructions, performs a handshake, and resets all semaphores to zero before proceeding (\Cref{sec:bg-barrier-based}).
This is reliable, but barriers serialize execution even when no cross-iteration dependencies exist, preventing overlap across iterations and leaving hardware idle while faster engines wait for the slowest.


We propose a \emph{barrier-free semaphore allocation} algorithm that replaces these barriers with precise, per-dependency wait conditions.
It operates on structured control-flow graphs (SCFGs) with arbitrarily nested, dynamically bounded loops and conditionals.
When all engines traverse a common SCFG, each engine tracks the iteration counts of its enclosing loops in local registers and computes each wait value in closed form from those counts and the dependency structure.
Each engine therefore stalls only for its actual dependencies---never for unrelated instructions on other engines.


This performance boost comes with risk: an incorrect semaphore assignment or wait-value computation causes silent data corruption or deadlock, and such bugs are timing-dependent, making them hard to reproduce.
We therefore start by defining when a dependency is \emph{satisfied}---the precise moment at which a consumer can safely issue without over-synchronizing.
We claim our semaphore allocation targets exactly this moment, allowing each instruction to issue if and only if the dependency is satisfied.
We formally prove this claim in the Lean proof assistant via \emph{bisimulation}~\cite{bisim} in approximately 15K lines with zero admitted lemmas.
The proof handles loop-carried dependencies and nested control flow, and is stated for our per-loop allocation, where multiple instructions in the same loop body share a single semaphore.


We implement both a barrier-based baseline and our new barrier-free algorithm as compiler backend passes at the AWS Neuron ISA level, each taking a kernel plus a dependency graph and emitting a kernel with semaphores allocated.
We work at the ISA level and hold the instruction schedule fixed, isolating the standalone effect of the allocation strategy.
%
On a microbenchmark isolating the barrier's cost, the barrier-free allocation achieves a $3.3\times$ speedup over the barrier-based baseline, and across a suite of ML kernels compiled from production NKI kernels~\cite{nkilib}---tiled matrix multiplication, RMSNorm, FlashAttention, AdamW, and Dropless MoE---we observe latency reductions of 10--45\%, matching or exceeding hand-tuned manual allocation on most kernels.


To the best of our knowledge, this work makes the following novel contributions:
\begin{itemize}
    \item A precise characterization of dependency satisfaction on multi-engine architectures---the minimum synchronization required for correctness (\Cref{sec:deps}).
    \item A barrier-free semaphore allocation algorithm that achieves this minimum and supports arbitrarily nested, dynamically bounded loops (\Cref{sec:algorithm}), together with an analysis of the resource tradeoffs across semaphore-allocation granularities (\Cref{sec:optimizations}).
    \item A machine-checked bisimulation proof in Lean for the allocation ($\sim$15K lines, 0 admitted lemmas) (\Cref{sec:verification}).
    \item An evaluation on Trainium showing a $3.3\times$ speedup on a microbenchmark isolating the barrier's cost and 10--45\% latency reductions across a suite of ML kernels relative to a barrier-based baseline (\Cref{sec:evaluation}).
\end{itemize}

\section{Motivating Example}
\label{sec:motivation}
\begin{figure}[!htbp]
\centering

\begin{tikzpicture}[
    font=\tiny,
    engine label/.style={font=\scriptsize\sffamily, anchor=east},
    title style/.style={font=\scriptsize\sffamily\bfseries, anchor=south west},
    code line/.style={font=\ttfamily\footnotesize, anchor=base west, inner sep=0pt},
    i0box/.style={draw, fill=orange!50, minimum height=2.6mm, inner sep=0pt, anchor=west, font=\scriptsize\sffamily, text=white},
    i1box/.style={draw, fill=blue!45,   minimum height=2.6mm, inner sep=0pt, anchor=west, font=\scriptsize\sffamily, text=white},
    i2box/.style={draw, fill=green!55!black, minimum height=2.6mm, inner sep=0pt, anchor=west, font=\scriptsize\sffamily, text=white},
    i3box/.style={draw, fill=red!55,    minimum height=2.6mm, inner sep=0pt, anchor=west, font=\scriptsize\sffamily, text=white},
    i4box/.style={draw, fill=purple!55, minimum height=2.6mm, inner sep=0pt, anchor=west, font=\scriptsize\sffamily, text=white},
    i5box/.style={draw, fill=teal!65,   minimum height=2.6mm, inner sep=0pt, anchor=west, font=\scriptsize\sffamily, text=white},
    barrier/.style={dashed, thick, gray!70},
    depedge/.style={-{Stealth[length=1.2mm]}, thick, red!80!black},
    looparrow/.style={-{Stealth[length=1.2mm]}, thick, black!80},
]

\begin{scope}
  \node[code line] (l1) at (0, 0)      {[Engine 0] $I_0$};
  \node[code line] (l2) at (0, -0.35)  {[Engine 1] $I_1$};
  \node[code line] (l3) at (0, -0.7)  {\textbf{loop} ...:};
  \node[code line] (l4) at (0.25, -1.05) {[Engine 2] };
  \node[code line] (i3) at ([xshift=0pt]l4.base east) {$I_2$};
  \node[code line] (l5) at (0.25, -1.40) {[Engine 1] };
  \node[code line] (i4) at ([xshift=0pt]l5.base east) {$I_3$};
  \node[code line] (l6) at (0, -1.75)  {[Engine 2] $I_4$};
  \node[code line] (l7) at (0, -2.1)  {[Engine 0] $I_5$};
  \draw[looparrow] (i3.east) .. controls +(0.35,0) and +(0.35,0) .. (i4.east);
  \node[font=\scriptsize, anchor=west] at ($(i4.east)+(0.18,0.33)$) {Loop-carry (2)};
\end{scope}

\begin{scope}[shift={(0.45\columnwidth, -0.2cm)}]
  \node[title style] at (1.0, -0.9) {(a) Barrier-based};
  
  \node[engine label] at (-0.1, 0.45) {Engine 0};
  \node[engine label] at (-0.1, 0.15) {Engine 1};
  \node[engine label] at (-0.1, -0.15) {Engine 2};

  \node[i0box, minimum width=10mm] at (0.00, 0.45) {$I_0$};
  \node[i5box, minimum width=10mm] at (3.40, 0.45) {$I_5$};

  \node[i1box, minimum width=5mm] at (0.00, 0.15) {$I_1$};
  \node[i3box, minimum width=8mm] at (1, 0.15) {$I_3$};
  \node[i3box, minimum width=8mm] at (1.8, 0.15) {$I_3$};
  \node[i3box, minimum width=8mm] at (2.6, 0.15) {$I_3$};
  \node[i2box, minimum width=5mm] at (1, -0.15) {$I_2$};
  \node[i2box, minimum width=5mm] at (1.8, -0.15) {$I_2$};
  \node[i2box, minimum width=5mm] at (2.6, -0.15) {$I_2$};
  \node[i4box, minimum width=5mm] at (3.4, -0.15) {$I_4$};

  \draw[barrier] (1, 0.7) -- (1, -0.40);
  \draw[barrier] (1.80, 0.7) -- (1.80, -0.40);
  \draw[barrier] (2.6, 0.7) -- (2.60, -0.40);
  \draw[barrier] (3.40, 0.7) -- (3.40, -0.40);
\end{scope}

\begin{scope}[shift={(0.45\columnwidth, -1.75cm)}]
  \node[title style] at (1, -0.9) {(b) Barrier-free};
  
  \node[engine label] at (-0.1, 0.45) {Engine 0};
  \node[engine label] at (-0.1, 0.15) {Engine 1};
  \node[engine label] at (-0.1, -0.15) {Engine 2};

  \node[i0box, minimum width=10mm] at (0.00, 0.45) {$I_0$};
  \node[i5box, minimum width=10mm] at (1, 0.45) {$I_5$};

  \node[i1box, minimum width=5mm] at (0.00, 0.15) {$I_1$};
  \node[i3box, minimum width=8mm] (e2i4a) at (0.5, 0.15) {$I_3$};
  \node[i3box, minimum width=8mm] at (1.30, 0.15) {$I_3$};
  \node[i3box, minimum width=8mm] at (2.1, 0.15) {$I_3$};

  \node[i2box, minimum width=5mm] at (0.00, -0.15) {$I_2$};
  \node[i2box, minimum width=5mm] at (0.5, -0.15) {$I_2$};
  \node[i2box, minimum width=5mm] (e3i3c) at (1.3, -0.15) {$I_2$};
  \node[i4box, minimum width=5mm] at (1.80, -0.15) {$I_4$};

  \draw[depedge, line width=1pt] (e3i3c.west) -- ($(e2i4a.south)!0.65!(e2i4a.south east)$);
\end{scope}

\end{tikzpicture}
\caption{\textbf{Motivating example.} A kernel with three engines and one cross-engine loop-carried dependency (offset~2). Dashed vertical lines in (a) mark all-engine barriers at every iteration boundary; in (b), a single red edge enforces the true dependency, letting all other engine work overlap.}
\label{fig:motivating-example}
\end{figure}
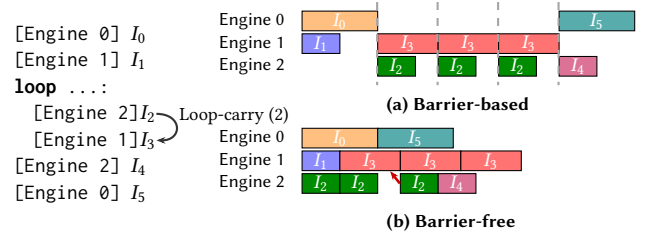

\Cref{fig:motivating-example} shows a small kernel that runs on three engines and highlights the cost of barrier-based synchronization.
Each line of the code assigns an instruction to an engine---for example, $I_0$ runs on Engine~0.
The loop body runs for three iterations in this example, and its body contains two instructions on different engines with a single inter-engine loop-carried dependency: $I_2$ (on Engine~2) in iteration $i$ waits for $I_3$ (on Engine~1) from iteration $i{-}2$.
The timeline omits control-related instructions (e.g., branches) and shows one possible execution in which each engine executes one instruction at a time in program order.
%

Using the barrier-based approach (\Cref{fig:motivating-example}(a)), the loop entry and every iteration boundary have all-engine barriers, shown as dashed vertical lines.
Only one dependency links two engines across an iteration distance of two, yet every iteration boundary forces all three engines to drain and handshake.
The pre-loop and post-loop work on Engine~0 is pushed outside the loop region, and Engine~0 sits idle for the entire loop.
%
Each iteration in the loop, Engine~2 waits for Engine~1 due to the barrier---even if Engine~2 is idle.

The barrier-free allocation (\Cref{fig:motivating-example}(b)) enforces only the single real dependency---the one red arrow, from the first completion of the producer $I_3$ to the third issue of the consumer $I_2$.
All other engine activity overlaps freely: the pre-loop and post-loop work on Engine~0 runs back-to-back while the loop executes on Engine~1 and Engine~2, and the loop iterations overlap across Engine~1 and Engine~2.
The speedup grows with loop trip count, since every eliminated barrier is overhead the barrier-based approach paid unconditionally.

\section{Background}
\label{sec:background}

We now review the Trainium architecture: multiple compute engines, instruction pipelines, and semaphore primitives that both the barrier-based and barrier-free approaches build on.

\subsection{Trainium Multi-Engine Architecture}
\label{sec:bg-architecture}

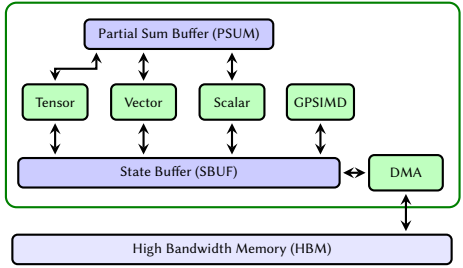
\begin{figure}[b]
\centering
\begin{tikzpicture}[
    engine/.style={draw, rounded corners=2pt, fill=green!25, minimum width=1.1cm, minimum height=0.6cm, font=\scriptsize\sffamily, align=center, thick},
    buffer/.style={draw, rounded corners=2pt, fill=blue!20, minimum height=0.5cm, font=\scriptsize\sffamily, align=center, thick},
    memory/.style={draw, rounded corners=2pt, fill=blue!10, minimum height=0.5cm, font=\scriptsize\sffamily, align=center, thick},
    arr/.style={<->, thick, >=stealth, shorten >=1pt, shorten <=1pt},
    outerbox/.style={draw=green!50!black, thick, rounded corners=4pt},
    scale=0.78, every node/.style={transform shape}
]

\node[buffer, minimum width=3.2cm] (psum) at (2.1, 0) {Partial Sum Buffer (PSUM)};

\node[engine] (tensor) at (0, -1.15) {Tensor};
\node[engine] (vector) at (1.5, -1.15) {Vector};
\node[engine] (scalar) at (3.0, -1.15) {Scalar};
\node[engine] (gpsimd) at (4.5, -1.15) {GPSIMD};

\node[buffer, minimum width=5.45cm] (sbuf) at (2.10, -2.35) {State Buffer (SBUF)};

\node[engine, minimum width=1.25cm] (dma) at (5.95, -2.35) {DMA};

\node[memory, minimum width=7.45cm] (hbm) at (3.00, -3.65) {High Bandwidth Memory (HBM)};

\node[outerbox, fit=(psum)(tensor)(gpsimd)(dma), inner sep=6pt] {};


\draw[arr] (tensor.north) -- ++(0, 0.25) -| ([xshift=-1.4cm]psum.south);
\draw[arr] (vector.north) -- ++(0, 0.15) -| ([xshift=-0.6cm]psum.south);
\draw[arr] (scalar.north) -- ++(0, 0.15) -| ([xshift=0.9cm]psum.south);

\draw[arr] (tensor.south) -- (tensor.south |- sbuf.north);
\draw[arr] (vector.south) -- (vector.south |- sbuf.north);
\draw[arr] (scalar.south) -- (scalar.south |- sbuf.north);
\draw[arr] (gpsimd.south) -- (gpsimd.south |- sbuf.north);

\draw[arr] (sbuf.east) -- (dma.west);

\draw[arr] (dma.south) -- (dma.south |- hbm.north);

\end{tikzpicture}
\caption{Trainium's NeuronCore architecture. Four compute engines share on-chip buffers SBUF and PSUM, while DMA moves data between SBUF/PSUM and HBM.}
\label{fig:neuroncore}
\end{figure}
\Cref{fig:neuroncore} illustrates the NeuronCore, the fundamental compute unit of Trainium~\cite{nki-arch}.
It comprises heterogeneous compute engines (e.g., tensor, vector, and scalar engines) interacting with the on-chip SRAM buffers (SBUF and PSUM) and a DMA engine for accessing High Bandwidth Memory (HBM)~\cite{trn1,trn2,trn3}. The different engines can work asynchronously on independent parts of a kernel at once, provided the data dependencies between them are correctly enforced.

\subsection{Instruction Types and Execution Pipelines}
\label{sec:bg-isa}

In the Trainium architecture~\cite{nki-arch}, each engine's instruction stream consists of two interleaved instruction types: \emph{datapath} instructions and \emph{control} instructions.
datapath instructions are the computational and memory operations (tensor contractions, vector or scalar arithmetic, and DMA transfers).
Each engine is \emph{deeply pipelined}: multiple datapath instructions per engine can be in-flight at once and their executions can be overlapped.
Datapath instructions issue to this pipeline and retire in-order.
Control instructions handle control flow (branches, loops) and register manipulation (arithmetic on registers).
Registers are engine-local, whereas semaphores (\Cref{sec:bg-semaphores}) are globally accessible across engines.
They execute atomically and are interleaved with datapath instructions in the instruction stream.
We make the precise ordering assumptions of this execution model explicit when we formalize it in \Cref{sec:verification}.
\begin{figure}[!htbp]
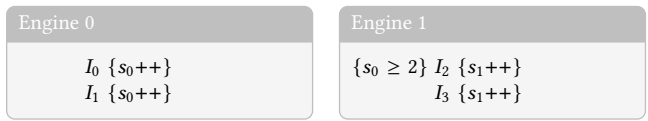

\centering
\begin{minipage}[t]{0.48\columnwidth}
\begin{tcolorbox}[colback=gray!8, colframe=gray!50, boxrule=0.5pt, left=0.5mm, right=0.5mm, top=0.5mm, bottom=0.5mm, title={\footnotesize Engine 0}]
\footnotesize
\begin{tabular}{@{\hspace{8mm}}r@{\;}l@{\;}l@{}}
              & \texttt{$I_0$} & $\{s_0\mathord{+}\mathord{+}\}$ \\
              & \texttt{$I_1$} & $\{s_0\mathord{+}\mathord{+}\}$ \\
\end{tabular}
\end{tcolorbox}
\end{minipage}%
\hfill
\begin{minipage}[t]{0.48\columnwidth}
\begin{tcolorbox}[colback=gray!8, colframe=gray!50, boxrule=0.5pt, left=0.5mm, right=0.5mm, top=0.5mm, bottom=0.5mm, title={\footnotesize Engine 1}]
\footnotesize
\begin{tabular}{@{}r@{\;}l@{\;}l@{}}
$\{s_0 \geq 2\}$ & \texttt{$I_2$} & $\{s_1\mathord{+}\mathord{+}\}$ \\
                 & \texttt{$I_3$} & $\{s_1\mathord{+}\mathord{+}\}$ \\
\end{tabular}
\end{tcolorbox}
\end{minipage}
\caption{\textbf{Straight-line semaphore allocation.} Each instruction increments its engine's semaphore on retirement. $I_2$ depends on $I_1$: the wait $s_0 \geq 2$ blocks $I_2$ until $I_1$ retires.}
\label{fig:straight-line-sema}
\end{figure}

\subsection{Semaphore-Based Synchronization}
\label{sec:bg-semaphores}
Trainium provides a synchronization primitive that implements a \emph{sema\-phore} as a mechanism to enforce dependencies.
A~semaphore is a shared 32-bit primitive that supports two operations: a \emph{wait}, which blocks an instruction's issue until the counter satisfies a comparison against a value, and an \emph{update}, which atomically modifies the counter when an instruction retires.

How the underlying \emph{dependency graph} is obtained is orthogonal to our work, which takes the graph as input and encodes it with semaphores.
The conventional encoding scheme uses a wait condition (\texttt{sem >= threshold}) and an increment-at-retirement (\texttt{sem += inc}), so each semaphore acts as a monotone counter of retirements.
The threshold can be an immediate constant or a register, enabling dynamically computed wait conditions.

Encoding a dependency is simple for straight-line code---the consumer waits for the producer to have retired by setting the \texttt{threshold} to the producer's ordinal position within its engine's instruction stream.
\Cref{fig:straight-line-sema} shows the code where one semaphore is allocated to each engine and each instruction increments its engine's semaphore on retirement.
Engine~0's semaphore $s_0$ counts its retirements, so the dependency where $I_2$ depends on $I_1$ is enforced by the wait $\mathtt{s_0 \geq 2}$, since $I_1$ is the second instruction in Engine~0's instruction stream. 
Thus, $I_2$ issues only once $I_1$ retires.
%

This scheme breaks down with loops: the wait value must account for \emph{which iteration} the producer and consumer are in.
We precisely specify \emph{dependency satisfaction} in \Cref{sec:deps}.

\subsection{A Barrier-Based Approach}
\label{sec:bg-barrier-based}

\begin{figure}[!htbp]
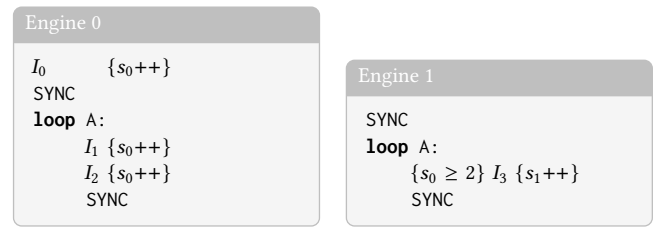

\centering
\begin{minipage}[t]{0.48\columnwidth}
\begin{tcolorbox}[colback=gray!8, colframe=gray!50, boxrule=0.5pt, left=0.5mm, right=0.5mm, top=0.5mm, bottom=0.5mm, title={\footnotesize Engine 0}]
\footnotesize
\begin{tabular}{@{}r@{\;}l@{\;}l@{}}
                 & $I_0$                       & $\{s_0\mathord{+}\mathord{+}\}$ \\
\multicolumn{3}{@{\hspace{1mm}}l}{\texttt{SYNC}} \\
\multicolumn{3}{@{\hspace{1mm}}l}{\texttt{\textbf{loop} A:}} \\
                 & \hspace{7mm}$I_1$          & $\{s_0\mathord{+}\mathord{+}\}$ \\
                 & \hspace{7mm}$I_2$          & $\{s_0\mathord{+}\mathord{+}\}$ \\
\multicolumn{3}{@{}l}{\hspace{8mm}\texttt{SYNC}} \\
\end{tabular}
\end{tcolorbox}
\end{minipage}%
\hfill
\begin{minipage}[t]{0.48\columnwidth}
\begin{tcolorbox}[colback=gray!8, colframe=gray!50, boxrule=0.5pt, left=0.5mm, right=0.5mm, top=0.5mm, bottom=0.5mm, title={\footnotesize Engine 1}]
\footnotesize
\begin{tabular}{@{\hspace{1mm}}r@{\;}l@{\;}l@{}}
\multicolumn{3}{@{\hspace{1mm}}l}{\texttt{SYNC}} \\
\multicolumn{3}{@{\hspace{1mm}}l}{\texttt{\textbf{loop} A:}} \\
\hspace{5mm} $\{s_0 \geq 2\}$ & $I_3$          & $\{s_1\mathord{+}\mathord{+}\}$ \\
\multicolumn{3}{@{}l}{\hspace{7mm}\texttt{SYNC}} \\
\end{tabular}
\end{tcolorbox}
\end{minipage}
\caption{\textbf{Barrier-based synchronization} for a same-iteration dependency where $I_3$ depends on $I_2$. Because \texttt{SYNC} drains and resets all semaphores at every iteration boundary, the wait threshold $s_0 \geq 2$ can be computed statically.}
\label{fig:barrier-based-example}
\end{figure}

When a loop's trip count is statically known it can be unrolled, restoring the straight-line scheme of \Cref{sec:bg-semaphores}; but when the trip count is determined at runtime, unrolling is no longer possible.
One approach that handles both cases is to insert an \emph{all-engine barrier} (\texttt{SYNC}) at every loop entry and iteration boundary.
Each barrier drains every engine's in-flight instructions, performs a cross-engine handshake, and resets all semaphores to zero, so each iteration again uses statically computed wait values (\Cref{fig:barrier-based-example}).

While easy to show correct, this approach has two fundamental limitations.
First, it \emph{over-synchronizes}: every engine waits at every barrier, even those with no dependency in that iteration, so no engine can run ahead into the next iteration or post-loop instructions.
Second, the drain, handshake, and reset are themselves costly, and this cost is paid once per iteration.
This motivates our barrier-free approach, which replaces these coarse-grained barriers with precise, per-dependency wait conditions that allow full engine parallelization.

\section{Defining Dependency Satisfaction}
\label{sec:deps}

We define when a dependency is \emph{satisfied}: the earliest point at which the producer has completed enough times for the consumer to issue.
This is the \emph{minimum synchronization required for correctness}: issue earlier and the consumer may execute before the producer, violating the dependency; later and the consumer waits longer than necessary.

\subsection{Program and Execution Model}
\label{sec:deps-scopes}

We target programs with \emph{structured control flow}, modeling every construct as a loop: an ordinary loop runs its trip count,  conditional branches run 0 or 1 times, and the program is a top-level loop that runs once.
This yields a tree of nested loops, where each non-top-level loop has a direct enclosing parent loop.
Following classical loop-nest dependency theory~\cite{allen2001optimizing}, an instruction $i$ is indexed by its \emph{iteration vector} $\vec{v}_i$: the iteration counts of its enclosing loops, outermost first.


\begin{figure*}[ht]
\centering
\begin{minipage}[c]{0.34\textwidth}
\begin{minipage}[t]{0.49\linewidth}
\begin{tcolorbox}[colback=gray!8, colframe=gray!50, boxrule=0.5pt, left=0.5mm, right=0.5mm, top=0.5mm, bottom=0.5mm, title={\footnotesize Engine 0}]
\begin{lstlisting}[style=pseudosmall, numbers=none]
loop A:
 loop B:
  Producer
  # (a,b)
\end{lstlisting}
\end{tcolorbox}
\end{minipage}%
\hfill
\begin{minipage}[t]{0.49\linewidth}
\begin{tcolorbox}[colback=gray!8, colframe=gray!50, boxrule=0.5pt, left=0.5mm, right=0.5mm, top=0.5mm, bottom=0.5mm, title={\footnotesize Engine 1}]
\begin{lstlisting}[style=pseudosmall, numbers=none]
loop A:
 loop B:
  Consumer
  # (a,b-2)
\end{lstlisting}
\end{tcolorbox}
\end{minipage}
\end{minipage}%
\hfill
\begin{minipage}[c]{0.62\textwidth}
\centering
\footnotesize
\begin{tabular}{@{}c c c l l@{}}
\hline
$\vec{v}_\cns$ & $\delta$ & $\vec{t}$ & $\sum_{\vec{i}\preceq\vec{t}} H_e(\vec{i}, \texttt{B})$ & Satisfaction \\
\hline
$(1,1)$ & $(0,2)$ & $(1,\text{-}1)$ & underflow & no wait \\
$(1,2)$ & $(0,2)$ & $(1,0)$ & underflow & no wait \\
$(1,3)$ & $(0,2)$ & $(1,1)$ & $\{(1{,}1)\} = 1$ & $\rc(\prd) \geq 1$ \\
$(2,1)$ & $(0,2)$ & $(2,\text{-}1)$ & underflow & no wait \\
$(2,2)$ & $(0,2)$ & $(2,0)$ & underflow & no wait \\
$(2,3)$ & $(0,2)$ & $(2,1)$ & $\{(1{,}1),(1{,}2),(1{,}3),(2{,}1)\} = 4$ & $\rc(\prd) \geq 4$ \\
$(2,4)$ & $(0,2)$ & $(2,2)$ & $\{(1{,}1),(1{,}2),(1{,}3),(2{,}1),(2{,}2)\} = 5$ & $\rc(\prd) \geq 5$ \\
\hline
\end{tabular}
\end{minipage}
\caption{\textbf{Loop-carried dependency, offset 2 on \texttt{B}} ($\delta = (0,2)$, $\parscope = \texttt{B}$). \texttt{B} runs 3 times under $\texttt{A}{=}1$ and 4 under $\texttt{A}{=}2$. Each \texttt{B}-entry contributes $H_e(\vec{i}, \texttt{B}) = 1$, so the sum counts iterations $\vec{i} \preceq \vec{t}$. A target with a non-positive component underflows, and the dependency is satisfied with no wait.}
\label{fig:dep-general}
\end{figure*}

\subsection{The Dependency-Satisfaction Condition}
\label{sec:model-deps}

A dependency makes a consumer $\cns$ wait for a producer $\prd$.
%
%
Following classical loop-nest theory~\cite{allen2001optimizing}, we describe the dependency between $\prd$ and $\cns$ by a \emph{distance vector} $\delta$ over their common loops: one component per common loop giving the iteration distance on that loop.
If the consumer's current iteration vector is $\vec{v}_\cns$, the producer iteration it depends on is the \emph{target} $\vec{t} = \vec{v}_\cns - \delta$.

$\vec{t} = \vec{v}_{\cns} - \delta$ is only well-defined when the producer and consumer are interpreted over a \emph{common structured control-flow graph}: both engines run the same loop nest, so $\delta$'s components refer to loops shared by producer and consumer.
The trip counts of those loops may still differ across engines.
We therefore assume a common SCFG throughout the rest of the paper and formalize it in \Cref{sec:verif-spec}.

Whether the dependency is satisfied is decided from the target $\vec{t}$ and two quantities: the producer's retirement count and the loop-entry counts.

\paragraph{Producer retirements}
$\rc(\prd)$ is the total number of times $\prd$ has retired; it is monotone.
Since $\prd$ appears once in the body of its \emph{parent loop} $\parscope$ (the loop immediately enclosing it) and retires once per issue (\Cref{sec:verif-spec}), $\rc(\prd)$ equals the number of completed entries of $\parscope$.

\paragraph{Loop-entry counts}
\label{sec:deps-runtime}
$H_e(\vec{i}, \ell)$ is the number of times engine $e$ enters loop $\ell$ during the common-loop iteration $\vec{i}$, where $\ell$ lies at or below the innermost loop $\vec{i}$ names.
This count is 1 when $\ell$ is the innermost common loop, and is equal to $\ell$'s trip count when $\ell$ is nested below it.
%

The consumer may issue once $\prd$ has retired in every common-loop iteration up to the target $\vec{t}$.
The target \emph{underflows}, written $\exists j:\ t_j \leq 0$, when some component is non-positive: it names an iteration before the first iteration of that loop, so no producer iteration is owed and the consumer issues with no wait.
Otherwise it waits for the cumulative producer executions up to $\vec{t}$.
The dependency is thus \emph{satisfied} if and only if
{\small
\begin{equation}
\label{eq:dep-sat}
\underbrace{\exists j:\ t_j \leq 0}_{\text{no-wait}}
\;\;\lor\;\;
\underbrace{\rc(\prd) \;\geq\; \sum_{\vec{i}\,\preceq\,\vec{t}} H_e(\vec{i}, \parscope)}_{\text{cumulative producer retirements through } \vec{t}}.
\end{equation}
}
Here $\preceq$ orders common-loop iterations lexicographically from the outermost loop inward, and each $\vec{i}$ contributes $H_e(\vec{i}, \parscope)$, the number of producer executions in that iteration.

\Cref{fig:dep-general} traces the condition on a loop-carried dependency.
Producer and consumer both sit in the inner loop \texttt{B} of \texttt{A} (so $\parscope = \texttt{B}$), with offset 2 on \texttt{B}: the consumer at $(a,b)$ reads what the producer wrote at $(a,b{-}2)$, giving $\delta = (0,2)$.
\texttt{B} runs 3 times under $\texttt{A}{=}1$ and 4 under $\texttt{A}{=}2$, so its trip count is not fixed.
Since the producer has no loop nested below \texttt{B}, it executes once per \texttt{B}-iteration, so $H_e(\vec{i}, \texttt{B}) = 1$ for every $\vec{i}$ and the sum simply counts iterations up to $\vec{t}$ (if the producer nested deeper, each term could exceed 1).
For the consumer at $(2,4)$, the target is $\vec{t} = (2,2)$ and the sum materializes as
{\small
\[
\begin{aligned}
\sum_{\vec{i}\,\preceq\,(2,2)} H_e(\vec{i}, \texttt{B})
&= \underbrace{H_e((1,1))+H_e((1,2))+H_e((1,3))}_{\texttt{A}{=}1:\ 3} \\
&\quad + \underbrace{H_e((2,1))+H_e((2,2))}_{\texttt{A}{=}2\ \text{through}\ b{=}2:\ 2}
\;=\; 5,
\end{aligned}
\]}%
so the consumer at $(2,4)$ may issue once the producer has retired 5 times.
\section{The Allocation Algorithm}
\label{sec:algorithm}
We now compile the dependency-satisfaction condition in \Cref{eq:dep-sat} into semaphore waits and updates, achieving the minimum synchronization for correctness.
Two architecture facts shape the compilation: semaphores are globally accessible while registers are engine-local (\Cref{sec:background}), and each instruction carries at most one semaphore update, applied at retirement.
We restrict to a practical subset of dependencies whose check reduces to a comparison against two registers and one shared semaphore, computed at issue time.
%
We present the intuition for the resulting closed-form wait values here; their exact correspondence to the dependency-satisfaction condition is established in the machine-checked proof of \Cref{sec:verification}.

\subsection{From the Condition to Registers}
\label{sec:alg-checks}

The condition in \Cref{eq:dep-sat} is a disjunction of two checks: a \emph{no-wait} check $\exists j: t_j \leq 0$, and a \emph{cumulative-retirement} check $\rc(\prd) \geq \sum_{\vec{i} \preceq \vec{t}} H_e(\vec{i}, \parscope)$.
The producer's retirement count $\rc(\prd)$ is the one cross-engine quantity in the condition, so we carry it in a semaphore, which is globally accessible and atomically incremented when the producer retires (\Cref{sec:bg-semaphores}).
The thresholds it is compared against are loop-entry counts that every engine tracks locally, so we hold those in registers.

Computing these thresholds at runtime is what costs registers: each is built from the per-iteration entry counts $H_e(\vec{i}, \ell)$, which an engine tracks by incrementing a register as it enters loops.
One loop's entry count is one register, but $H$ is indexed by the full iteration vector---an inner loop's count depends on which iteration of every enclosing loop it falls in.
This means that to support all possible dependencies for a loop nested under $d$ enclosing loops, we must track a separate entry count for each combination of the enclosing loops' iterations, growing exponentially in $d$, so materializing $H$ in registers is infeasible.


Instead, we restrict to two per-loop counters and confine the distance vector's only nonzero component to the innermost common loop between producer and consumer.
Reducing dependency structure to a scalar distance on a shared loop is common in loop optimizations: modulo scheduling, for instance, models each loop-carried dependency by a scalar distance on the scheduled loop~\cite{oppermann2019modulosched}.

Let $\shared$ be the innermost common loop of $\prd$ and $\cns$, and $\parscope$ the producer's parent loop---either $\shared$ itself or a loop nested inside it.
We require the dependency to be \emph{carried at $\shared$}: $\prd$ and $\cns$ agree on every common loop strictly above $\shared$, so the distance vector has a single nonzero component there, the \emph{offset} $k \geq 0$.
The gap is now one scalar, and each check reduces to a register comparison.

\textbf{No-wait check via trip register.} The \emph{trip register} $\hat{r}_\ell$ incremented at every entry of $\ell$ but reset to 0 whenever $\ell$'s parent is re-entered, so it counts the entries of $\ell$ in the current trip through its parent.
The only level whose target can underflow is $\shared$, so the no-wait check becomes a single comparison $\hat{r}_\shared \leq k$.
It holds when at most $k$ iterations of $\shared$ have run in the current trip: the offset then points before this trip began, so no producer is owed.
When every dependency carried at $\shared$ has offset 0, the no-wait check $\hat{r}_\shared \leq k$ becomes $\hat{r}_\shared \leq 0$, which is false whenever the consumer issues (the loop has been entered, so $\hat{r}_\shared \geq 1$), so we omit $\hat{r}_\shared$ for that loop.

\textbf{Cumulative retirements via monotone register.} The \emph{monotone register} $r_\ell$ incremented at every entry of $\ell$ and never reset, counting the total entries of $\ell$.
%


\subsection{The Allocatable Cases}
\label{sec:alg-registers}

Carried at $\shared$, the cumulative count depends only on where the producer sits relative to $\shared$.
The dependency is \emph{forward}~\cite{allen2001optimizing} if $\cns$ follows $\prd$ in $\shared$'s body and \emph{backward} otherwise.
If $\prd$ sits inside a loop $B$ nested in $\shared$, we use $B$'s position in $\shared$'s body as $\prd$'s (likewise for $\cns$).
Because $\prd$ and $\cns$ may run on different engines, this order is forced only when a nested loop or block boundary separates them; when they share the same straight-line region of $\shared$ from different engines, we take the dependency to be forward.
This leaves three cases, by where $\prd$ sits relative to $\shared$ and the order of the dependency.
%
%

\begin{figure}
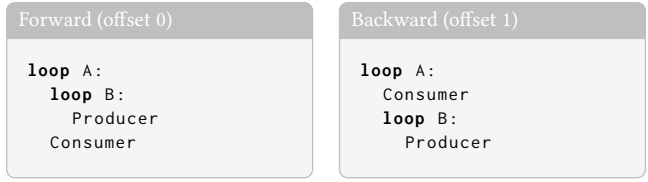

\centering
\begin{minipage}[t]{0.48\columnwidth}
\begin{tcolorbox}[colback=gray!8, colframe=gray!50, boxrule=0.5pt, left=0.5mm, right=0.5mm, top=0.5mm, bottom=0.5mm, title={\footnotesize Forward (offset 0)}]
\begin{lstlisting}[style=pseudosmall, numbers=none]
loop A:
  loop B:
    Producer
  Consumer
\end{lstlisting}
\end{tcolorbox}
\end{minipage}%
\hfill
\begin{minipage}[t]{0.48\columnwidth}
\begin{tcolorbox}[colback=gray!8, colframe=gray!50, boxrule=0.5pt, left=0.5mm, right=0.5mm, top=0.5mm, bottom=0.5mm, title={\footnotesize Backward (offset 1)}]
\begin{lstlisting}[style=pseudosmall, numbers=none]
loop A:
  Consumer
  loop B:
    Producer
\end{lstlisting}
\end{tcolorbox}
\end{minipage}
\caption{\textbf{Producer nested below $\shared$.} Both share $\shared = A$ and producer parent $\parscope = B$, with $\cns$ directly in $A$'s body. Producer and consumer may run on different engines; only their relative nesting and order in $A$'s body matter here, so we use a single stream. Forward: $B$ precedes $\cns$, so $r_B$ already counts $B$ through the current $A$-iteration. Backward: $\cns$ precedes $B$, so $r_B$ counts $B$ through $A$'s previous iteration. Either way the wait value is $r_B$.}
\label{fig:dep-nested}
\end{figure}

\begin{itemize}[nosep, leftmargin=1.5em]
\item \textbf{Producer in $\shared$ ($\parscope = \shared$), forward or backward}: $\prd$ retires once per iteration of $\shared$, so the count is $r_\shared - k = r_\parscope - k$, for any offset $k \geq 0$ (\Cref{fig:dep-general}).
\item \textbf{Producer nested, forward} (offset 0): $\prd$ is in a loop $\parscope$ nested inside $\shared$, before $\cns$ in $\shared$'s body; when $\cns$ issues, its engine has finished this iteration's trip through $\parscope$, so the count is $r_\parscope$ (\Cref{fig:dep-nested} left).
\item \textbf{Producer nested, backward} (offset 1): $\cns$ precedes $\prd$'s loop, so its engine has not entered $\parscope$ this iteration; the count through $\shared$'s previous iteration is again $r_\parscope$ (\Cref{fig:dep-nested} right).
\end{itemize}

The nested cases admit only these two offsets: offset 0 when forward and offset 1 when backward.
A backward dependency at offset 0 is unsatisfiable, as the consumer would wait on a producer iteration that has not yet run.
Any larger offset requires additional registers to recover how many times $\parscope$ executed in earlier trips through $\shared$; the monotone counter $r_\parscope$ only records total entries, not the number of entries for a specific past iteration of $\shared$ (e.g., B's total iteration count varies in different trips through A in \Cref{fig:dep-general}).

Once the no-wait pre-check fails, the wait value is in every case a formula in the single monotone register $r_\parscope$: we write $\mathit{expr}$ for it, equal to $r_\parscope - k$ for a producer in $\shared$ and $r_\parscope$ for the two nested cases.

Dependencies with a multi-level gap or any other nested offset fall back to the barrier-based scheme (\Cref{sec:bg-barrier-based}); our evaluation (\Cref{sec:evaluation}) shows the subset suffices for our target AI kernels.

\subsection{A Single Branchless Wait Value}
\label{sec:alg-masking}

The register-level check is a disjunction: the no-wait pre-check $\hat{r}_\shared \leq k$, or the cumulative check against the wait value $\mathit{expr}$ of \Cref{sec:alg-registers}.
A semaphore wait, however, takes a single threshold computed into a wait register before issue, not a branch.
We collapse the disjunction with a \emph{multiplicative mask}: a boolean materialized as a $0/1$ integer and multiplied into the threshold.
We use an $\isGT$ operation with $\isGT(a,b) = 1$ iff $a > b$.
We compute
\begin{equation}
\label{eq:wait-val}
\mathit{mask} \;\triangleq\; \isGT(\hat{r}_\shared,\, k),
\qquad
\mathit{waitVal} \;\triangleq\; \mathit{mask} \cdot \mathit{expr}.
\end{equation}
The mask is the complement of the no-wait condition $\hat{r}_\shared \leq k$: it is 0 exactly when no producer is owed, forcing $\mathit{waitVal} = 0$ so any non-negative semaphore passes; otherwise it is 1 and the threshold is $\mathit{expr}$.
The single check $\rc(\prd) \geq \mathit{waitVal}$ is thus equivalent to the full disjunction in \Cref{eq:dep-sat}, emitted as a fixed three-operation sequence (compute mask, compute $\mathit{expr}$, multiply) before each consumer issue.

\subsection{Per-Instruction Allocation}
\label{sec:alg-per-instr}

\begin{figure}[t]
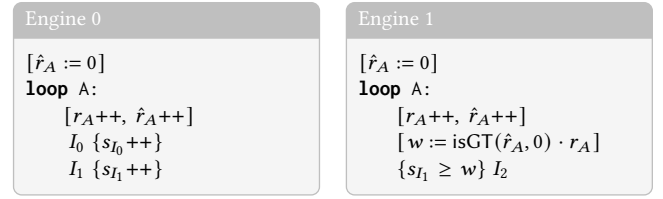

\centering
\begin{minipage}[t]{0.48\columnwidth}
\begin{tcolorbox}[colback=gray!8, colframe=gray!50, boxrule=0.5pt, left=0.5mm, right=0.5mm, top=0.5mm, bottom=0.5mm, title={\footnotesize Engine 0}]
\footnotesize
\begin{tabular}{@{}r@{\;}l@{\;}l@{}}
\multicolumn{3}{@{}l}{$[\hat{r}_A := 0]$} \\
\multicolumn{3}{@{}l}{\texttt{\textbf{loop} A:}} \\
\multicolumn{3}{@{}l}{\hspace{5mm}$[r_A\mathord{+}\mathord{+},\ \hat{r}_A\mathord{+}\mathord{+}]$} \\
 & \hspace{5mm}$I_0$ & $\{s_{I_0}\mathord{+}\mathord{+}\}$ \\
 & \hspace{5mm}$I_1$ & $\{s_{I_1}\mathord{+}\mathord{+}\}$ \\
\end{tabular}
\end{tcolorbox}
\end{minipage}%
\hfill
\begin{minipage}[t]{0.48\columnwidth}
\begin{tcolorbox}[colback=gray!8, colframe=gray!50, boxrule=0.5pt, left=0.5mm, right=0.5mm, top=0.5mm, bottom=0.5mm, title={\footnotesize Engine 1}]
\footnotesize
\begin{tabular}{@{}r@{\;}l@{\;}l@{}}
\multicolumn{3}{@{}l}{$[\hat{r}_A := 0]$} \\
\multicolumn{3}{@{}l}{\texttt{\textbf{loop} A:}} \\
\multicolumn{3}{@{}l}{\hspace{5mm}$[r_A\mathord{+}\mathord{+},\ \hat{r}_A\mathord{+}\mathord{+}]$} \\
\multicolumn{3}{@{}l}{\hspace{5mm}$[w := \isGT(\hat{r}_A, 0)\cdot r_A]$} \\
\hspace{5mm}$\{s_{I_1} \geq w\}$ & $I_2$ & \\
\end{tabular}
\end{tcolorbox}
\end{minipage}
\caption{\textbf{Per-instruction allocation.} Each instruction gets its own semaphore, incremented at retirement; the bracketed \texttt{regOp}s maintain the monotone register $r_A$ and trip register~$\hat{r}_A$. Consumer $I_2$ waits on the producer's semaphore $s_{I_1}$ with the masked wait value.}
\label{fig:per-instr-example}
\end{figure}

It remains to supply $\rc(\prd)$ through a semaphore.
The simplest scheme is \emph{per-instruction}: assign one semaphore $s_\prd$ to each producer, incremented when $\prd$ retires, so $s_\prd = \rc(\prd)$ exactly.
The dependency check is
\begin{equation}
\label{eq:per-instr-check}
s_\prd \;\geq\; \isGT(\hat{r}_\shared,\, k) \cdot \mathit{expr},
\end{equation}
with the consumer computing $\mathit{waitVal}$ using three control-arithmetic operations before waiting on $s_\prd$.
\Cref{fig:per-instr-example} shows an inter-engine dependency $I_2$ depending on $I_1$, with registers and wait value computations in place to realize the per-instruction allocation.

A production workload can comprise millions of instructions, so one semaphore per producer does not scale.
This motivates our new allocation algorithm.

\subsection{Per-Loop Allocation}
\label{sec:alg-per-scope}

The per-loop allocation, our main algorithm (\Cref{sec:verification}), assigns one semaphore per (loop, engine) pair, shared by all instructions in that loop body on that engine.
A shared counter no longer equals a single producer's retirement count, but we recover it in closed form, since all instructions sharing a parent loop execute the same number of times on every trace.

Within a loop body holding $N$ instructions on a given engine, each increments the loop's shared semaphore $s_{loop}$ (in place of the per-producer $s_\prd$ above) by 1 at retirement.
After the $K$-th of them (by position in the body) retires in the $r$-th iteration, $s_{loop} = N \cdot r - (N - K)$, so a producer $\prd$ at position $K$ has retired at least $r$ times exactly when
\begin{equation}
\label{eq:per-loop-count}
\mathit{s_{loop}} \;\geq\; N \cdot r - (N - K).
\end{equation}

\begin{figure}[t]
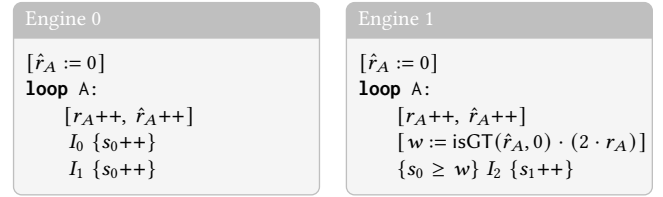

\centering
\begin{minipage}[t]{0.48\columnwidth}
\begin{tcolorbox}[colback=gray!8, colframe=gray!50, boxrule=0.5pt, left=0.5mm, right=0.5mm, top=0.5mm, bottom=0.5mm, title={\footnotesize Engine 0}]
\footnotesize
\begin{tabular}{@{}r@{\;}l@{\;}l@{}}
\multicolumn{3}{@{}l}{$[\hat{r}_A := 0]$} \\
\multicolumn{3}{@{}l}{\texttt{\textbf{loop} A:}} \\
\multicolumn{3}{@{}l}{\hspace{5mm}$[r_A\mathord{+}\mathord{+},\ \hat{r}_A\mathord{+}\mathord{+}]$} \\
 & \hspace{5mm}$I_0$ & $\{s_0\mathord{+}\mathord{+}\}$ \\
 & \hspace{5mm}$I_1$ & $\{s_0\mathord{+}\mathord{+}\}$ \\
\end{tabular}
\end{tcolorbox}
\end{minipage}%
\hfill
\begin{minipage}[t]{0.48\columnwidth}
\begin{tcolorbox}[colback=gray!8, colframe=gray!50, boxrule=0.5pt, left=0.5mm, right=0.5mm, top=0.5mm, bottom=0.5mm, title={\footnotesize Engine 1}]
\footnotesize
\begin{tabular}{@{}r@{\;}l@{\;}l@{}}
\multicolumn{3}{@{}l}{$[\hat{r}_A := 0]$} \\
\multicolumn{3}{@{}l}{\texttt{\textbf{loop} A:}} \\
\multicolumn{3}{@{}l}{\hspace{5mm}$[r_A\mathord{+}\mathord{+},\ \hat{r}_A\mathord{+}\mathord{+}]$} \\
\multicolumn{3}{@{}l}{\hspace{5mm}$[w := \isGT(\hat{r}_A, 0)\cdot(2\cdot r_A)]$} \\
\hspace{5mm}$\{s_0 \geq w\}$ & $I_2$ & $\{s_1\mathord{+}\mathord{+}\}$ \\
\end{tabular}
\end{tcolorbox}
\end{minipage}
\caption{\textbf{Per-loop allocation} for the program of \Cref{fig:per-instr-example}. Engine~0's two instructions now share one semaphore $s_0$, so the wait value scales by $N{=}2$ and offsets the producer's position $K$.}
\label{fig:per-scope-example-alloc}
\end{figure}

This holds because in-order retirement (\Cref{sec:verif-spec}), combined with sequential issue within a body, keeps all retirement counts in the loop within 1 of each other---the first few at $m{+}1$, the rest at $m$.
Substituting the masked wait value in \Cref{eq:wait-val} for $r$ in \Cref{eq:per-loop-count} gives the per-loop check:

\begin{equation}
\label{eq:per-loop-check}
s_{loop} \;\geq\; \isGT(\hat{r}_\shared,\, k) \cdot \bigl(N \cdot \mathit{expr} - (N - K)\bigr),
\end{equation}
where $s$ is the semaphore of $\prd$'s loop on $\prd$'s engine, and $N$, $K$ are static.
This adds two operations to the per-instruction sequence---the $N\cdot{}$ scaling and the $-(N-K)$ offset---for five in total per issue.
\Cref{fig:per-scope-example-alloc} shows the allocation on the same program from \Cref{fig:per-instr-example}.

\section{Resource Tradeoffs}
\label{sec:optimizations}


Although the wait-value computation of \Cref{sec:algorithm} is fixed, we can trade the number of semaphores against per-issue ALU work by choosing how many producers share a semaphore.
A~more widely shared semaphore must be \emph{decoded} to recover a single producer's retirement count $\rc(\prd)$.

%
%
%
We make this tradeoff explicit, adding a third \emph{per-engine} point to the two allocations of \Cref{sec:algorithm}, and then show how standard compiler passes cut the ALU cost.
The per-engine allocation and the ALU-reduction passes are unverified extensions on top of the per-loop algorithm we verify in \Cref{sec:verification}.

\subsection{The Semaphore--ALU Tradeoff}
\label{sec:opt-per-engine}

Each allocation maps a producer to one semaphore and recovers $\rc(\prd)$ from its value by a decoding whose cost grows as the semaphore is shared more widely.
\Cref{sec:algorithm} gave two: per-instruction \Cref{eq:per-instr-check}, with one semaphore per producer ($s_\prd = \rc(\prd)$, no decoding, $O(N)$ semaphores), and per-loop \Cref{eq:per-loop-check}, with one semaphore per (loop, engine) pair, decoded by $N \cdot \mathit{expr} - (N - K)$ ($O(L+\cdot E)$ semaphores).
We now add the coarsest point.

\textbf{Per-engine.}
Since in-order retirement holds across all of an engine's instructions (\Cref{sec:verification}), one counter per engine tracks its total retirements---folding in everything that retires before $\prd$ on that engine, not just $\prd$'s own loop.

Recovering $\rc(\prd)$ therefore means accounting for every retirement before $\prd$ on its engine, summing the loop-entry counts $H$ over all loops on the path from program entry to $\prd$.
The per-loop closed form $N \cdot \mathit{expr} - (N - K)$ in \Cref{eq:per-loop-check} applies to each such loop $l$, reducing to the full body count $N_l(r_l - 1)$ for the other loops that executed before $\prd$ and to the precise $N_\parscope \cdot \mathit{expr} - (N_\parscope - K)$ for $\prd$'s parent loop $\parscope$; summing over the path gives $\rc(\prd)$.
The no-wait condition is handled as before, by the same mask $\isGT(\hat{r}_\shared, k)$ in \Cref{eq:wait-val} applied to the entire summed wait value.

This uses only $O(E)$ semaphores but $O(D)$ ALU operations per issue, where $D$ is the number of loops on that path.

Each step down in semaphore count thus pushes more of the recovery of $\rc(\prd)$ onto runtime ALU work.

\subsection{Reducing Per-Issue ALU Cost}
\label{sec:opt-passes}

Standard compiler passes reduce the per-issue ALU cost without affecting correctness.

\textbf{Loop-invariant code motion.} Each loop's contribution to the sum changes only when that loop is re-entered, so it is invariant to any loop nested deeper.
Hoisting it out of those deeper loops leaves, in steady state, only the innermost contribution plus a read of a cached outer partial sum, at the cost of a few registers per engine to hold the cached sums.

\textbf{Common-subexpression elimination.} Consumers sharing enclosing loops can share most of the wait-value compu\-tation---the register reads, the masking $\isGT$, and the partial sums up to their closest common ancestor---so reusing one consumer's computation across the others amortizes the cost.


\subsection{Resource Summary}
\label{sec:opt-summary}

\Cref{tab:alloc-resources} summarizes the three allocations, differing only in resource cost.
A compiler can mix them across a program---per-loop where the dependency footprint is small, per-engine for deeply nested regions under a tight semaphore budget---trading semaphores, registers, and ALU work according to their relative cost on the target.

\begin{table}[t]
\centering
\caption{Resource cost across the three allocations, before the ALU reductions of \Cref{sec:opt-passes}. $N$ is the number of datapath instructions, $L$ the number of loops, $E$ the number of engines, and $D$ the number of loops on the path to the producer.}
\label{tab:alloc-resources}
\small
\begin{tabular}{@{}lccc@{}}
\toprule
Granularity & Semaphores & Registers/engine & ALU/issue \\
\midrule
Per-instruction & $O(N)$        & $O(L)$ & 3 \\
Per-loop        & $O(L\cdot E)$ & $O(L)$ & 5 \\
Per-engine      & $O(E)$        & $O(L)$ & $O(D)$ \\
\bottomrule
\end{tabular}
\end{table}
\section{Verified Correctness}
\label{sec:verification}

We formalize the multi-engine execution model of \Cref{sec:deps} as a \emph{specification} transition system and the register-and-counter execution model of \Cref{sec:algorithm} as an \emph{implementation} transition system.
The implementation abstracts the functionality of a semaphore as a \emph{shared counter}: a global counter that any engine may increment and read. The results therefore apply to any architecture providing such counters.
We then prove via \emph{bisimulation}~\cite{bisim} that, whenever an implementation program realizes a specification program via our per-loop allocation of \Cref{sec:alg-per-scope}, every wait fires if and only if its dependency is satisfied without under- or over-synchronization.
%

Bridging the two requires showing $r_\ell$ counts the total entries of loop $\ell$, $\hat{r}_\ell$ its entries in the current trip, and the register check in \Cref{eq:per-loop-check} coincides with the dependency satisfaction condition in \Cref{eq:dep-sat}. This holds for programs related by our allocation, over our restricted set of dependencies.

The proof is mechanized in Lean in approximately 15K lines with zero admitted lemmas.

\subsection{Specification}
\label{sec:verif-spec}

We first formalize the semantic dependency-satisfaction model of \Cref{sec:deps} as a small-step transition system.

\paragraph{Abstracting away the ISA}

We aim to reason about execution traces and the possible relative orderings of instructions; concrete instruction semantics are irrelevant to the proof.
We therefore abstract datapath instructions opaquely and keep only their effect on a shared \emph{datapath state}, consisting of HBM, SBUF, PSUM, and the datapath registers.
Engines are likewise treated abstractly, while the \emph{control state} is engine-local and consists only of the control registers used to steer structured execution.

\paragraph{Program}
We inductively define statements as:
\[
\mathit{Stmt}\ S ::= \mathbf{block}(f) \mid \mathbf{loop}(g, u, \overline{S}) \mid \mathbf{cond}(g, u_T, u_F, \overline{S}_T, \overline{S}_F)
\]

A specification \emph{program} $P_s$ consists of a $Stmt$ list and a dependency graph. The grammar keeps a separate $\mathbf{cond}$ form for faithfulness to the definition of SCFG, but as in \Cref{sec:deps-scopes} a conditional is just a loop of trip count 0 or 1, so we refer only to loops hereafter. Here, $f$ maps each engine to its datapath instruction list for that block, $g$ is a guard function from engines' control state to a Boolean, the $u$-functions update control state upon exiting loops, and $\overline{S}$ is a nested statement list.
The program's $Stmt$ structure defines the SCFG common across engines, though the instruction streams and dynamic control decisions may differ by engine.
The proof therefore assumes no agreement on control decisions or trip counts across engines; \Cref{sec:divergence} explains why the bisimulation holds in such cases.


\paragraph{State}
A specification state is a tuple $\sigma_s = (\mathit{ds}, \mathit{cs}, \kappa, \mathit{ifl}, H, \rc)$
consisting of the shared datapath state $\mathit{ds}$, a map $\mathit{cs}$ from engines to their individual control states, a per-engine continuation stack $\kappa$ (a path from the root down to the currently executing loop and the progress made within each loop), a per-engine in-flight list $\mathit{ifl}$ of instructions in the pipeline, the loop-entry counts $H$ of \Cref{sec:deps-runtime}, and the retirement counts $\rc$.
\appdetail{The precise state typing is given in \Cref{sec:appendix-verif-spec-state}.}

\paragraph{Transition system}
The step relation fires on a single non-deterministically chosen engine per step, so the specification admits every engine interleaving consistent with the dependency-satisfaction condition in \Cref{eq:dep-sat}.
Datapath instructions progress through an issue/commit/retire lifecycle: they issue only when their dependency is satisfied and retire in order, incrementing $\rc$ at retirement.
We model overlap among in-flight instructions by allowing a non-deterministically chosen instruction in the pipeline to commit its update to the shared datapath state.
Separate control-flow steps evaluate guards, perform loop entry and exit, update the continuation stack, and maintain $H$ for use in the dependency predicate. \appdetail{The detailed step rules are given in \Cref{sec:appendix-verif-spec-steps}.}


\subsection{Implementation}
\label{sec:verif-impl}

We next formalize a concrete counter-and-register execution model.
This model is expressive enough to realize the masked wait-value computation and counter updates of \Cref{sec:alg-registers,sec:alg-masking} needed to instantiate a per-loop allocation.

\paragraph{Program}
An implementation program $P_i$ has the same statement grammar as a specification program, but block streams may additionally contain inserted $\mathbf{regOp}$ statements and there is no dependency graph.
Instead, registers compute wait values while the program also records explicit wait and signal counter annotations for each datapath instruction.
%

\paragraph{State}
The implementation state $\sigma_i$ is defined similarly to $\sigma_s$ as a tuple $\sigma_i = (\mathit{ds}, \mathit{cs}, \kappa, \mathit{ifl}, \mathit{sema}, \mathit{regs})$. Note that the shared counter state and the engine-local register files are in place of the maps $H$ and $R$.

\paragraph{Transition system}
The implementation transition system augments the specification's with $\mathbf{regOp}$ steps, checks of the shared counter at issue time, and counter updates at retirement.
\appdetail{The rules are given in \Cref{sec:appendix-verif-impl-steps}.}

\subsection{Allocation Relation}
\label{sec:verif-alloc}

To prove the correctness of the algorithm independent of a particular compiler implementation, inspired by CompCert~\cite{compcert}, we define an \emph{allocation relation} $\mathcal{R}_{\mathrm{alloc}}$: a declarative predicate on $(P_s, P_i)$ program pairs stating that $P_i$ realizes $P_s$ via our per-loop allocation.

The relation says that the two programs agree on the opaque instruction semantics and engine set, and that the implementation has the same statement-tree structure as the specification save for inserted $\mathbf{regOp}$ statements in block streams.
In particular, the inserted $\mathbf{regOp}$ statements properly maintain $r$ and $\hat{r}$ for each loop, shared counters are injectively assigned per (loop, engine) pair, and wait/update annotations are consistent with the per-loop decoding of \Cref{sec:alg-per-scope}.

\subsection{Proof of Correctness}
\label{sec:verif-what}

Our correctness criterion is \emph{bisimulation}~\cite{bisim} between the specification and implementation: a simulation relation $(\sim)$ on states that holds at corresponding initial states and is preserved by every step on either side, where the responding side may take zero or more steps of its own to match.

\paragraph{Simulation relation}
A simulation relation is a binary relation between specification and implementation states.
In our proof, the relation is defined by
\[
\sigma_s \sim \sigma_i \;\triangleq\; \mathit{MatchStates}(\sigma_s, \sigma_i) \,\wedge\, \mathit{SpecInv}(\sigma_s) \,\wedge\, \mathit{ImplInv}(\sigma_i).
\]

$\mathit{MatchStates}(\sigma_s, \sigma_i)$ asserts the cross-system invariants: datapath-state equality, per-engine in-flight queue equality, control-state equality, and continuation-stack correspondence modulo inserted $\mathbf{regOp}$ statements.
It also relates the registers to $H$: for every loop $\ell$ and engine $e$, $r_\ell$ holds the total entries of $\ell$ and $\hat{r}_\ell$ its entries in the current trip.
$\mathit{SpecInv}$ and $\mathit{ImplInv}$ are the side-specific invariants the proof maintains on each system.

We prove bisimulation under the following hypotheses, which we unpack before stating the final theorem.

\begin{itemize}[nosep, leftmargin=1.5em]
\item $\mathcal{R}_{\mathrm{alloc}}(P_s, P_i)$ of \Cref{sec:verif-alloc}, stating that $P_i$ realizes $P_s$ via our per-loop allocation.
\item $\mathsf{Allocatable}(P_s)$---every dependency in $P_s$ falls into one of the allocatable cases of \Cref{sec:alg-registers}.
\item $\mathsf{UniqueLoopIds}(P_s)$ and $\mathsf{UniqueInstrIds}(P_s)$, so that identifiers assigned to instructions and loops are unambiguous keys into the $H$ and $\rc$ maps.
\end{itemize}

\begin{theorem}[Bisimulation]
\label{thm:cr-bisimulation}
Let $P_s$ and $P_i$ satisfy the four hypotheses above.
Then from any pair of common initial states $\sigma_{s_0}$ and $\sigma_{i_0}$\appdetail{ (\Cref{sec:appendix-verif-init})}, the specification $P_s$ and implementation $P_i$ are bisimilar under $\sim$. That is,

\begin{enumerate}
\item $\sigma_{s_0} \sim \sigma_{i_0}$;
\item for every $\sigma_s \sim \sigma_i$ and every specification state $\sigma_s'$ reached by one step from $\sigma_s$, the implementation can take steps from $\sigma_i$ to some state $\sigma_i'$ such that $\sigma_s' \sim \sigma_i'$;
\item for every $\sigma_s \sim \sigma_i$ and every implementation state $\sigma_i'$ reached by one step from $\sigma_i$, the specification can take steps from $\sigma_s$ to some state $\sigma_s'$ such that $\sigma_s' \sim \sigma_i'$.
\end{enumerate}
\end{theorem}

%

\subsection{Bisimulation Under Control-Flow Divergence}
\label{sec:divergence}
Under the common SCFG assumption, loop \emph{trip counts} may still differ across engines (\Cref{sec:verif-spec}), so a target $\vec{t}$ can name a producer iteration that never runs---leaving the dependency permanently blocked.

Concretely, consider a dependency with offset 0 from a producer on one engine to a consumer on another, both in a loop that runs zero times on the producer's engine but once on the consumer's.
This does not break the bisimulation: on the specification side the producer's retirement count never reaches 1, and on the implementation side the producer's shared counter never reaches the expected wait value of 1. Thus, neither the specification nor the implementation can take the issue step for the consumer.

Under the hypotheses of \Cref{thm:cr-bisimulation}, the two systems block together in such cases, so their states remain related.
Liveness is therefore relative to the specification induced by the input dependency graph: the implementation introduces no \emph{additional} blocking, but faithfully reproduces any deadlock already present in the specification.


\subsection{Consequences}

Because ${\sim}$ is preserved on both sides and includes equality of the shared datapath state at every related pair, \Cref{thm:cr-bisimulation} yields trace and reachable-state equivalence: the specification and implementation admit exactly the same set of datapath traces.
It also yields \emph{safety} (the implementation never issues before a dependency is satisfied) and \emph{liveness} (when the specification can progress, so can the implementation), so it never over-synchronizes relative to the input dependency graph.


\subsection{Trusted Computing Base}
\label{sec:verif-tcb}

\Cref{thm:cr-bisimulation} holds modulo trusted assumptions.
First, the specification and implementation (\Cref{sec:verif-spec,sec:verif-impl}) faithfully model multi-engine execution on real hardware; in particular, the datapath, control, and allocation registers are disjoint---so computation never clobbers the counters and wait values the $\mathbf{regOp}$s maintain---and the compiler is trusted to allocate the three separately.
Second, $\mathcal{R}_{\mathrm{alloc}}$ faithfully reflects the allocation of \Cref{sec:alg-per-scope}.
Third, the syntactic hypotheses on $P_s$ hold ($\mathsf{Allocatable}$, $\mathsf{UniqueLoopIds}$, $\mathsf{UniqueInstrIds}$).
Finally, the Lean kernel is trusted, as is standard.

%
%
%
%
\section{Evaluation}
\label{sec:evaluation}

We evaluate the barrier-free semaphore allocation algorithm on Trainium hardware around four questions:

\par{\textbf{RQ1 (Latency vs.\ barriers):} how much latency does barrier-free synchronization save over barrier-based?

\par{\textbf{RQ2 (vs.\ unrolling):}} how does it compare to loop unrolling in latency and code size?

\par{\textbf{RQ3 (vs.\ manual):}} how competitive is the barrier-free allocation with hand-tuned expert placement?

\par{\textbf{RQ4 (Resource usage):}} what is the allocation's register and semaphore overhead? Does it match the bounds of \Cref{sec:optimizations}?}



\subsection{Experimental Setup}
\label{sec:eval-setup}

\paragraph{Hardware}
We run all experiments on an AWS Trainium~2 chip through an Amazon EC2 \texttt{trn2.48xlarge} instance.
Each kernel runs on a single Neuron core.
Benchmarks are written in the AWS Neuron ISA for Trainium translated from production NKI kernels~\cite{nkilib}.

\paragraph{Variants}
We compare four variants of each kernel that enforce the same data dependencies but differ in how they synchronize them; the dependencies are manually extracted from the bare, unsynchronized ISA program:


\begin{itemize}[nosep, leftmargin=1.5em]
\item \textbf{Barrier-Based} --- the baseline: at every loop iteration boundary, all engines drain their in-flight instructions, perform an all-engine handshake, reset all semaphores, and handshake again before proceeding.
To isolate the effect of the synchronization strategy, we implement this scheme directly in our AWS Neuron ISA backend rather than compare against the production compiler.
\item \textbf{Barrier-Free} --- our verified per-loop allocation (\Cref{sec:alg-per-scope}).
\item \textbf{Manual} --- semaphore allocation placed manually by a developer familiar with the AWS Neuron ISA.
\item \textbf{Unrolled} --- the loop fully unrolled with statically computed semaphore waits; unavailable for Dropless MoE which has a dynamically bounded loop.
\end{itemize}

\paragraph{Benchmarks}
We evaluate one microbenchmark and a suite of ML kernels, chosen to span the patterns the algorithm targets: static and dynamically bounded loops, single/nested loops and a range of dependency structures. All dependency offsets fall within the allocatable cases of \Cref{sec:alg-registers}.
\begin{itemize}[nosep, leftmargin=1.5em]
\item \textbf{AddManyLoop} --- our microbenchmark: a $10^4$-iteration loop of element-wise tensor addition with no inter-iteration dependencies, isolating the pure barrier overhead.
\item \textbf{TiledMatmul} --- a tiled $2048^3$ matrix multiply over M, N, and K, with the DMA, Tensor, and Vector engines cooperating per tile. Its innermost K loop is a tight, compute-bound body of a weight load and a matmul. We evaluate it with all three loops as hardware loops, and in a \emph{K-unrolled} variant whose innermost loop is statically unrolled.
\item \textbf{RMSNorm} --- RMS normalization with FP8 quantization over a $32768 \times 8192$ input; software-pipelined, double-buffering loads on one engine against stores on another.
\item \textbf{FlashAttention} --- tiled attention ($8192^2 \times 128$) with two nested loops over query groups and key/value sections; heavy DMA-prefetch/Tensor-compute overlap.
\item \textbf{AdamW} --- the Adam optimizer with weight decay over 1M ($2^{20}$) parameters; a loop over parameter chunks with several engines cooperating per chunk.
\item \textbf{Dropless MoE} --- a Mixture-of-Experts MLP iterating over the \emph{active} expert blocks, a count set by the runtime token-to-expert routing not the input shape. We evaluate three configurations of increasing iteration count, varying the tokens $T$ and experts $E$ with top-$k$ routing fixed at $k{=}2$: $(T{=}384, E{=}4)$, $(T{=}2048, E{=}8)$, and $(T{=}4096, E{=}16)$.
\end{itemize}

\subsection{Results}
\label{sec:eval-latency}


\begin{table*}[t]
    \centering
    \caption{Four synchronization variants on Trainium~2 per benchmark. $L$ is the number of hardware loops; Offsets are the dependency offsets present in each benchmark; Norm.\ Lat is end-to-end latency normalized to the barrier-based baseline, so lower is better and $0.50$ means half the baseline latency; Code is code size; Regs is registers added per engine for synchronization; Sema is the number of semaphores. Unrolled is unavailable (N/A) for Dropless MoE kernels due to dynamically bounded loops.}
    \label{tab:alloc-sema-combined}
    \resizebox{\textwidth}{!}{%
    \begin{tabular}{lcc|rrrr|rrrr|rrrr|rrrr}
      \toprule
      & & & \multicolumn{4}{c|}{Barrier-Based} & \multicolumn{4}{c|}{Barrier-Free} & \multicolumn{4}{c|}{Manual} & \multicolumn{4}{c}{Unrolled} \\
      Benchmark & $L$ & Offsets & \makecell{Norm.\\Lat} & Code & Regs & Sema & \makecell{Norm.\\Lat} & Code & Regs & Sema & \makecell{Norm.\\Lat} & Code & Regs & Sema & \makecell{Norm.\\Lat} & Code & Regs & Sema \\
      \midrule
      AddManyLoop & 1 & $0$ & 1.00 & 2.5 KB & 0 & 3 & 0.30 & 1.7 KB & 4 & 4 & 0.22 & 1.8 KB & 0 & 4 & 0.04 & 1.2 MB & 0 & 4 \\
      AdamW 1M & 1 & $0,2$ & 1.00 & 5.3 KB & 0 & 4 & 0.74 & 7.3 KB & 4 & 10 & 0.76 & 5.3 KB & 5 & 4 & 0.76 & 11.8 KB & 0 & 4 \\
      flash\_attn 8192$^2$ & 2 & $0,1,2$ & 1.00 & 22.5 KB & 0 & 8 & 0.86 & 37.1 KB & 8 & 19 & 0.88 & 23.5 KB & 0 & 8 & 0.70 & 1.2 MB & 0 & 8 \\
      rmsnorm 32768x8192 & 1 & $0,2$ & 1.00 & 4.0 KB & 0 & 4 & 0.57 & 6.1 KB & 4 & 6 & 0.56 & 3.9 KB & 5 & 4 & 0.56 & 508.6 KB & 0 & 4 \\
      tiled\_matmul 2048$^3$ & 3 & $0,1,2$ & 1.00 & 4.7 KB & 0 & 5 & 0.90 & 6.3 KB & 11 & 7 & 0.71 & 3.7 KB & 7 & 4 & 0.39 & 177.5 KB & 0 & 4 \\
      tiled\_matmul 2048$^3$ (K-unrolled) & 2 & $0,1,2$ & 1.00 & 6.9 KB & 0 & 5 & 0.55 & 9.3 KB & 8 & 6 & 0.61 & 6.1 KB & 7 & 4 & 0.55 & 177.5 KB & 0 & 4 \\
      MoE 384 (E=4, k=2) & 1 & $0,1$ & 1.00 & 10.3 KB & 0 & 21 & 0.87 & 9.3 KB & 4 & 13 & 0.81 & 9.8 KB & 4 & 21 & N/A & N/A & N/A & N/A \\
      MoE 2048 (E=8, k=2) & 1 & $0,1$ & 1.00 & 10.3 KB & 0 & 21 & 0.88 & 9.3 KB & 4 & 13 & 0.79 & 6.1 KB & 4 & 21 & N/A & N/A & N/A & N/A \\
      MoE 4096 (E=16, k=2) & 1 & $0,1$ & 1.00 & 10.3 KB & 0 & 21 & 0.88 & 9.3 KB & 4 & 13 & 0.79 & 6.1 KB & 4 & 21 & N/A & N/A & N/A & N/A \\
      \bottomrule
    \end{tabular}%
    }
\end{table*}

\Cref{tab:alloc-sema-combined} reports four metrics per variant.
\emph{Lat} is end-to-end latency normalized to the barrier-based baseline (lower is better), and \emph{Code} is the code size.
\emph{Sema} is the number of semaphores the variant uses.
\emph{Regs} is the number of registers the allocation adds \emph{per engine} for synchronization, beyond those the original kernel uses for its own computation: the loop counters $r_\ell, \hat{r}_\ell$, the masks, and the temporaries that compute wait values.
Since the allocation inserts the same register families on every engine (\Cref{sec:alg-checks}), this number of registers is identical across engines.

First, note the robustness of the per-loop allocation with the allocatable subset of \Cref{sec:alg-registers}, successfully synchronizing every benchmark spanning dense nested loops (TiledMatmul, FlashAttention), software-pipelined loops (RMSNorm, AdamW), and dynamically routed workloads (Dropless MoE).

\paragraph{RQ1: latency versus barriers}
The barrier-free allocation improves over the barrier-based baseline on every benchmark.
%
The barrier-free allocation reduces latency by 70\% on AddManyLoop, whose independent iterations make every barrier pure overhead, and by 10--45\% on the real kernels, since the algorithm stalls only on each true dependency rather than on all engines.

\paragraph{RQ2: versus unrolling}
Unrolling a loop removes its per-iteration synchronization and branching, but produces code size proportional to the iteration count---up to three orders of magnitude larger here (RMSNorm 6.1\,KB vs.\ 508.6\,KB, FlashAttention 37.1\,KB vs.\ 1.2\,MB).
%
On most kernels barrier-free allocation comes within a few percent of the unrolled latency while keeping code compact (RMSNorm 0.57 vs.\ 0.56, AdamW 0.74 vs.\ 0.76).
The two TiledMatmul variants show where this breaks down.
TiledMatmul is compute-bound, and its innermost K loop is a tight body of two datapath instructions (a weight load and a matmul).
%
With all three loops kept, barrier-free must, every K iteration and on the same Tensor engine that runs the matmul, take the loop branch and update its monotone and trip counters---overhead that lands directly on the critical path and leaves it far behind unrolling (0.90 vs.\ 0.39).
%
Statically unrolling just that innermost loop moves the counter updates out to once per tile, where they hide behind the matmul, and barrier-free returns to par with unrolling (0.55 for both).
Barrier-free thus composes well with selective unrolling and delivers a larger gain where its wait-value ALU cost can be hidden behind other engines' work.
Note Dropless MoE cannot be unrolled at all---its loop runs over the \emph{active} expert blocks, set by the runtime token-to-expert routing rather than the input shape.

\paragraph{RQ3: versus manual expert placement}
%
%
%
%

On most kernels the barrier-free allocation is comparable to manual allocation, and sometimes beats it by precisely encoding the dependencies.
An expert may add conservative over-synchronization for safety.
%
%
%
The gap appears when the wait-value ALU cannot hide behind the engines' other work: sharing one semaphore per loop, the barrier-free algorithm decodes the semaphore to recover the producer's retirement count, and also maintains monotone and trip counters.

An expert can omit the decode step by giving the producer its own semaphore (i.e., selectively applying the per-instruction allocation) or comparing against a register the kernel already maintains (e.g., the native loop counter).
On TiledMatmul the decode and counter updates fall on the critical path (RQ2); the expert replaces them with direct register comparisons, using even fewer semaphores than barrier-free (4 vs.\ 7).
On Dropless MoE the expert instead takes the first route, spending more semaphores---21 vs.\ 13---to read each count directly.
It also dedicates a register to each wait threshold and updates them once per iteration at the loop tail, sharing them across all waits in the body, whereas the barrier-free allocation reuses a temporary and so recomputes a decode before every wait.

Making these choices well---which producers earn a dedicated semaphore or register, and how many the budget can afford---takes repeated profiling, critical-path inspection, and re-validation of the hand-placed semaphores on every change.
This is feasible for small kernels, but scales poorly to larger code while risking subtle correctness bugs.

\paragraph{RQ4: resource usage}
The Regs and Sema columns empirically track the bounds of \Cref{tab:alloc-resources}.
%
Registers stay within $O(L)$: 4 per engine for single-loop kernels, rising to 8 for the two-loop kernels and 11 for the three-loop kernel.
The ALU-reduction passes (\Cref{sec:opt-passes}) and the no-wait trip-register elision (\Cref{sec:alg-checks}) keep these counts low.
%
Semaphore counts scale with the loops and engines a dependency spans, ranging from 4--13, up to 19 on FlashAttention.
The barrier-based baseline adds no registers, trading all its synchronization cost into per-iteration barrier latency; the Manual variant's budget is comparable (0--7 per engine), trading semaphores against the runtime decode as RQ3 details.

%
%

%
%
%
%
%

\paragraph{Summary}
The barrier-free allocation beats the barrier-based baseline on every benchmark and, though fully automated, matches hand-tuned manual and unrolled performance on most kernels.
Unlike unrolling, it handles dynamically bounded loops and keeps code compact.
Its wait-value ALU mostly hides behind other engines' compute; the exception is a tight, compute-bound innermost loop (TiledMatmul), resolved by unrolling it while keeping the outer loops.
\section{Threats to Validity}
\label{sec:threats}


\paragraph{Coverage.} The study spans a microbenchmark and a suite of ML kernels---enough to show the dependencies we target arise in practice, but not an exhaustive sample of workloads.
Kernels with different engine usage or memory behavior may shift the magnitude of the speedups. 
The manual baseline is one reasonable expert allocation, not a latency lower bound.

\paragraph{Isolation.} We evaluate synchronization at the AWS Neuron ISA level, holding the instruction schedule fixed. Results report the standalone effect of the synchronization strategy rather than end-to-end compiler speedups.
Our contribution is orthogonal to instruction-scheduling techniques (e.g., software pipelining), which speed up the baseline and our allocation alike. Our work targets a different axis: ensuring synchronization does not block available cross-engine parallelism.

\paragraph{Dependency graphs.} The dependencies driving the barrier-free and manual variants are extracted by hand. Correctness is validated against a NumPy reference model.
The residual risk is the opposite of corruption: a conservative dependency graph leads both variants to over-synchronize.

\paragraph{Generality.} The gains require that the target is a multi-engine accelerator that synchronizes through semaphore-like, shared counters, and require workloads with recoverable cross-iteration overlap; on inherently serial loops, or hardware where ALU work is costly relative to global synchronization, the advantage shrinks.
Results may also shift across hardware generations and configurations, but the algorithm applies to any such multi-engine system.

\paragraph{Verification scope.} We verify the per-loop allocation under the execution model of \Cref{sec:verification}, modulo the trusted computing base (\Cref{sec:verif-tcb}). 
The optimizations of \Cref{sec:optimizations} are unverified.
\section{Related Work}
\label{sec:related}


\paragraph{Synchronization primitives in parallel architectures}

Synchronization primitives in parallel systems range from coarse-grained barriers~\cite{hensgen1988barrier} to fine-grained wait-free algorithms~\cite{herlihy1991waitfree}.
GPU programming relies on thread-block barriers and warp-level primitives in the SIMT model~\cite{cuda-programming-guide}.
Among ASIC accelerators, Groq's TSP is fully deterministic and software-scheduled by design, with the compiler planning all instruction timing statically~\cite{abts2020thinkfast,abts2022softwaredefinedtsp,groq2024predictability}; Google's TPUs~\cite{jouppi2017tpu,jouppi2023tpuv4,tpu-sysarch} exhibit heterogeneous compute units, but their published sources do not describe on-chip inter-unit synchronization.
By contrast, Trainium exposes a multi-engine model with explicit semaphore-based synchronization and dynamic control flow, so the compiler must decide when each engine may proceed.
DeNovo~\cite{denovo,denovosync} provides synchronization for disciplined parallelism at the hardware-protocol level, whereas our work operates at compiler-time allocation.

\paragraph{Synchronization reduction}

Reducing synchronization overhead is a long-studied problem.
%
%
In the HPC literature, Tseng~\cite{tseng1995barrier} introduced the closest ancestor to our approach---compiler analysis that eliminates SPMD barriers in favor of lighter-weight synchronization---and subsequent work extended barrier elimination via access-dependency analysis in OpenMP~\cite{openmp-barrier-elim}, dynamic runtime elision in production codes~\cite{chabbi2015barrier}, and point-to-point producer--consumer synchronization that signals specific consumers rather than all threads~\cite{p2p-sync}.
Xiao et al.~\cite{inter-block-gpu} proposed lock-free inter-block GPU synchronization.
These techniques target homogeneous SPMD or threaded programs on shared memory, whereas we eliminate barriers across heterogeneous engines that share no common barrier primitive, replacing them with point-to-point waits derived per engine.

\paragraph{Compiler frameworks and scheduling for AI accelerators}
To achieve high compute utilization for tensor workloads, many compiler frameworks separate the algorithm from its schedule~\cite{halide,tvm}, or the data placement from the computation placement for dataflow accelerators~\cite{hsu2025stardust}. Others provide multi-level IRs~\cite{mlir} or polyhedral loop optimization~\cite{pluto}, including polyhedral kernel generation for NPUs~\cite{zhao2021akg}.
A closer line of work schedules data movement and execution overlap directly: scheduling languages make GPU memory-hierarchy mapping explicit~\cite{hagedorn2020fireiron,hagedorn2023graphene}, and recent systems overlap computation across distributed kernels~\cite{zheng2025tritondist} or jointly schedule warp specialization for tensor-core kernels~\cite{soi2025twill}.
These techniques (e.g., software pipelining) structure execution to expose parallelism.
Our work is orthogonal: when parallelism is present, we generate fine-grained synchronization to achieve it.

\paragraph{Verified compiler passes}

Our bisimulation proof follows the simulation-based semantic-preservation methodology developed in CompCert~\cite{compcert}, which is used broadly in verified compilers, including CakeML~\cite{cakeml-backend}.
%
Schneider et al.~\cite{schneider2016inductive} prove simulation-based compiler correctness by induction on program structure.
Individual passes have also been verified, including instruction and superblock scheduling~\cite{yang2024scheduling,six2022superblock} and register allocation~\cite{verified-ra-framework,tristan2008validators}---a similar kind of resource-allocation problem to our semaphore allocation.
%
The verified Lustre compiler~\cite{lustre-verified}, also built on CompCert, instead targets a synchronous-reactive dataflow language whose logical lock-step clocks differ fundamentally from the asynchronous, semaphore-coordinated engines we compile for.
%
Like us, Burrow reasons about the space of allowed concurrent executions, but its framework targets weak-memory consistency mappings (e.g., x86 to Arm), whereas we prove preservation of synchronization across asynchronous engines~\cite{sprokholt2026burrow}.

\paragraph{Formal methods for AI accelerator compilation}

Formal methods are increasingly applied across the layers of the AI-compilation stack: graph-rewrite verification in XLA over unbounded tensor ranks~\cite{tensorright}, verified functional-tensor compilation~\cite{liu2024atl}, verification dialects embedded in MLIR~\cite{fehr2025verifdialects}, translation validation for Halide schedules~\cite{clement2022halidetv}, and equivalence checking of distributed training plans~\cite{trainverify}.
%
These efforts verify tensor-level computation, such as graph rewrites, schedules, and training plans, whereas we verify the hardware-specific code generation beneath them: the synchronization and parallel execution of an accelerator's engines.

\section{Conclusion}
\label{sec:conclusion}
%
%
%

We identified the exact synchronization condition for dependencies on a multi-engine AI accelerator, and compiled it into barrier-free, closed-form semaphore waits for arbitrarily nested loops with dynamic bounds.
We proved the per-loop allocation correct in Lean via bisimulation, and implemented it as a compiler backend pass at the Neuron ISA level.
%
On real kernels, it reduces latency by 10--45\% relative to the barrier-based baseline (and achieves a $3.3\times$ speedup on a synchronization-bound microbenchmark), remaining competitive with expert hand-tuning without the code-size blowup of unrolling.
The result is a fully automated synchronization scheme that is precise, mechanically verified, and free of both barriers and over-synchronization.

\bibliographystyle{ACM-Reference-Format}
\bibliography{refs}

\clearpage
\appendix

\section{Verification}
\label{sec:appendix-verification}
\noindent
The mechanization uses a map-based representation: every loop and every datapath instruction is assigned a unique identifier, written $\mathit{LoopId}$ and $\mathit{DataPathInstrId}$, which serve as the keys for the internal maps below. This is the content of the $\mathsf{UniqueLoopIds}$ and $\mathsf{UniqueInstrIds}$ hypotheses in \Cref{sec:verif-what}.

\subsection{Specification State}
\label{sec:appendix-verif-spec-state}

The specification state is
\[
\sigma_s = (\mathit{ds},\ \mathit{cs},\ \kappa,\ \mathit{ifl},\ H,\ \rc).
\]
Here $\mathit{ds}$ is the shared datapath state. The remaining components are maps:
\[
\begin{aligned}
\mathit{cs} &: \mathit{EngineId} \to \mathit{ControlState},\\
\kappa &: \mathit{EngineId} \to \mathit{ContStack},\\
\mathit{ifl} &: \mathit{EngineId} \to \mathit{InFlightList},
\end{aligned}
\]
\[
\begin{aligned}
H &: \mathit{EngineId} \to \mathit{LoopId} \to \mathbb{N} \to \mathit{LoopId} \to \mathbb{N},\\
\rc &: \mathit{DataPathInstrId} \to \mathbb{N}.
\end{aligned}
\]
In the proof code, dependencies are represented only as triples $(\prd,\cns,k)$ where the offset $k$ is on the shared loop $S$, so we index $H$ as the number of entries of loop $P$ during the $i_S$-th entry of $S$, where $i_S$ is $S$'s monotone entry count. This is equivalent to the iteration-vector presentation in the paper body for the shared-loop case, because once $S$ is fixed, each monotone entry of $S$ bijects with exactly one common-loop iteration vector at $S$.
The specification program components used by these rules are likewise treated abstractly but have fixed shape. For a block statement $\mathbf{block}(f)$, the block stream is
\[
\qquad
f : \mathit{EngineId} \to \mathit{DataPathInstrId}\ \mathit{list},
\]
while the shared functions appearing in loop/conditional constructors and the opaque instruction semantics are typed by
\[
\begin{aligned}
\instrOp &: \mathit{DataPathInstrId} \to \mathit{DataPathState}\\
          &\quad\to \mathit{DataPathState},\\
\guardFn &: \mathit{EngineId} \to \mathit{LoopId} \to \mathit{ControlState} \to \mathit{Bool},\\
\ctrlOp &: \mathit{EngineId} \to \mathit{LoopId} \to \mathit{ControlState} \\ &\quad\to \mathit{ControlState}.
\end{aligned}
\]
The dependency graph is likewise keyed by instruction identifiers, and this is the type-level information suppressed in the main text.

\subsection{Specification Step Rules}
\label{sec:appendix-verif-spec-steps}

The specification small-step relation chooses one engine non-deterministically at each step. The detailed rule families are below.
We write only loops here, though the proof code has separate entry and exit rules for loops and conditionals.

\begin{enumerate}
\item \textbf{Issue.} The next instruction in the current block may issue only if its dependency-satisfaction predicate in \Cref{eq:dep-sat} holds. The step appends the instruction to the in-flight list as \emph{issued} and advances the block position.
\item \textbf{Commit.} A non-deterministically chosen \emph{issued} in-flight instruction applies its opaque datapath semantics to the shared datapath state and is marked \emph{committed}. This is how the model represents overlap among in-flight instructions.
\item \textbf{Retire.} The head of the in-flight list, if already committed, is removed and the corresponding producer retirement count in $\rc$ is incremented. Retirement is strictly in issue order.
\item \textbf{Loop entry.} A loop guard is evaluated. On loop entry, a new continuation-stack frame is pushed and the loop-entry histories/counts used by the dependency predicate are updated.
\item \textbf{Loop exit / back edge.} When execution reaches the end of a loop body, the corresponding frame is popped; loops then return to their header for guard re-evaluation.
\item \textbf{Block completion.} When all instructions of a block have issued, the continuation stack advances to the next statement.
\end{enumerate}

\subsection{Implementation Program Components}
\label{sec:appendix-verif-impl-program}

An implementation program again consists of a structured statement tree, but each block stream may interleave datapath instructions with inserted $\mathbf{RegOp}$ instructions:
\[
f_i : \mathit{EngineId} \to (\mathit{DataPathInstrId} + \mathit{RegOp})\ \mathit{list}.
\]
It also carries two explicit semaphore-annotation maps on datapath instructions,
\[
\begin{aligned}
\mathit{waitOf} &: \mathit{DataPathInstrId} \to \mathit{SemaId},\\
\mathit{updateOf} &: \mathit{DataPathInstrId} \to \mathit{SemaId},
\end{aligned}
\]
identifying, respectively, which semaphore is checked before the \textbf{Issue} step and which semaphore is incremented at \textbf{Retire}.

\subsection{Implementation State}
\label{sec:appendix-verif-impl-state}

The implementation state modifies the specification state with shared semaphore state and engine-local register files. Concretely, the tuple is
\[
\sigma_i = (\mathit{ds},\ \mathit{cs},\ \kappa,\ \mathit{ifl},\ \mathit{sema},\ \mathit{regs}),
\]
where the new map components are
\[
\begin{aligned}
\mathit{sema} &: \mathit{SemaId} \to \mathbb{N},\\
\mathit{regs} &: \mathit{EngineId} \to \mathit{RegId} \to \mathbb{N}.
\end{aligned}
\]

Under the semaphore allocation, an implementation maintains one distinguished register on each engine serving as the wait register, holding the computed $\mathit{waitVal}$ read by issue-time semaphore checks.
The information $\rc$ and $H$ carry semantically is instead recovered operationally, through $\mathbf{RegOp}$ counter maintenance and \textbf{Issue}-/\textbf{Retire}-time semaphore effects.

%

\subsection{Implementation Step Rules}
\label{sec:appendix-verif-impl-steps}

The implementation follows the same structured control-flow skeleton as the specification, but with three additional operational features.
First, each inserted $\mathbf{RegOp}$ executes atomically as a control-side step: it reads the current engine-local register file, performs the prescribed ALU computation, and writes back the updated register values in one transition.
The mechanization supports the operations relevant to our proofs: addition, subtraction, and multiplication, each in immediate and register forms, together with $\mathsf{isGT}$; all are modeled with natural-number arithmetic.
Second, \textbf{Issue} no longer consults the semantic dependency predicate directly; instead, for a datapath instruction $I$, the implementation uses $\mathit{waitOf}(I)$ to select the semaphore associated with $I$ and checks that this semaphore is at least the value currently stored in the distinguished wait register.
Third, \textbf{Retire} uses $\mathit{updateOf}(I)$ to determine which semaphore to increment when a producer instruction retires.
These are the only new effects beyond the specification's behavior.



\subsection{Initial States}
\label{sec:appendix-verif-init}

The bisimulation theorem starts the specification and implementation from \emph{common initial states}: both systems are built from the same shared datapath state value $\mathit{ds}_0$ and the same per-engine control states $\mathit{cs}_0$, with all engines positioned at the first statement of the top-level loop.

Concretely, let $\kappa_0$ be the initial continuation-stack map pointing each engine to the top-level statement list at its first statement, with initial block offset $0$. Let $[\ ]_{\mathit{ifl}}$ map each engine to the empty in-flight list, and let $\mathbf{0}_{\rc}$, $\mathbf{0}_{H}$, $\mathbf{0}_{\mathit{sema}}$, and $\mathbf{0}_{\mathit{regs}}$ denote the corresponding constant-zero functions. The specification initial state is
\[
\sigma_{s_0} = (\mathit{ds}_0,\ \mathit{cs}_0,\ \kappa_0,\ [\ ]_{\mathit{ifl}},\ \mathbf{0}_{H},\ \mathbf{0}_{\rc}),
\]
and the implementation initial state is
\[
\sigma_{i_0} = (\mathit{ds}_0,\ \mathit{cs}_0,\ \kappa_0,\ [\ ]_{\mathit{ifl}},\ \mathbf{0}_{\mathit{sema}},\ \mathbf{0}_{\mathit{regs}}).
\]

Here, the empty in-flight lists and the zero maps indicate that no instruction has issued yet, no loop-entry or retirement counts have been accumulated, and all registers and semaphores start at zero.

\end{document}